# Latent class mixed modelling for phenotypic stratification of primary biliary cholangitis patients on first line treatment

Victoria Mulcahy[1,2*], Anais Rouanet[3*], Alessio Gerussi[4,5], Adam Duckworth[6], Steve Flack[1], Marco Carbone[4,5], Brian Tom[3], George Mells[1,2]

[1] Academic Department of Medical Genetics, University of Cambridge, Cambridge, UK

[2] Cambridge Liver Unit, Cambridge University Hospitals NHS Foundation Trust, Cambridge, UK

[3] MRC Biostatistics Unit, Cambridge Institute of Public Health, Forvie Site, Robinson Way, Cambridge Biomedical Campus, Cambridge CB2 0SR, UK

[4] Division of Gastroenterology, Centre for Autoimmune Liver Diseases, Department of Medicine and Surgery, University of Milano-Bicocca, Monza, Italy

[5] European Reference Network on Hepatological Diseases (ERN RARE-LIVER), San Gerardo Hospital, Monza, Italy

[6] Department of Histopathology, Cambridge University Hospitals NHS Foundation Trust

*joint first authors

**Financial support statement:** VM is funded by a Clinical Research Training Fellowship from the MRC, UK (MR/S021744/1). GFM was funded by a Post-doctoral Fellowship from the NIHR Rare Diseases – Translational Research Consortium (RD-TRC), UK, and is now funded by a Clinical Academic Research Partnership from the MRC, UK (MR/T023848/1). GFM has received funding support for UK-PBC from Intercept Pharmaceuticals.

* **Address for correspondence:**

George Mells

Academic Department of Medical Genetics

University of Cambridge

Cambridge, UK

Email: gfm26@medschl.cam.ac.uk

Phone: 01223 21611

**Abbreviations**

| | |
|---|---|
| ALP | Alkaline phosphatase |
| ALT | Alkaline transaminase |
| AIC | Akaike Information Criterion |
| AUC | Area under the curve |
| BIC | Bayesian Information Criterion |
| CA | Cholangitis activity |
| CI | Confidence interval |
| ESLD | End-stage liver disease |
| HA | Hepatitis activity |
| LCMM | Latent class mixed modelling |
| PBC | Primary biliary cholangitis |
| UDCA | Ursodeoxycholic acid |

**Lay summary**

Primary biliary cholangitis (PBC) is a chronic liver disease in which the small bile ducts inside the liver are injured by a person's own immune system. PBC is treated in the first instance with a medication called ursodeoxycholic acid (UDCA). Most patients respond well to treatment with UDCA (identified as UDCA responders), however, some fail to do so. This study focuses on designing ways to identify subgroups of patients with PBC (from those who respond well to treatment through to those who do not at all) which is vital to inform future studies aimed at understanding the causal differences between them. With this information, newer treatment options and improved patient outcomes in the long term are possible.

**Abstract**

**Background & Aims:** In patients with primary biliary cholangitis (PBC), the serum liver biochemistry measured during treatment with ursodeoxycholic acid (the UDCA response) accurately predicts long-term outcome. In this study we sought to use liver biochemistry, and in particular alkaline phosphatase (ALP), as a surrogate marker of disease activity, for phenotypic stratification in PBC using a computational modelling approach. Our aim here was to identify distinct disease subgroups of patients with distinct disease trajectories.

**Methods:** We used longitudinal ALP results from 1,601 PBC patients on first line treatment with UDCA, and applied latent class mixed modelling (LCMM), to identify distinct phenotypic subgroups, each with distinct disease trajectories, and risks of end stage liver disease (ESLD).

**Results:** We identified four well discriminated phenotypic subgroups within our PBC cohort, each with distinct disease trajectories.

**Introduction**

Primary biliary cholangitis (PBC) is a chronic autoimmune liver disease that is characterised by progressive cholestasis, biliary fibrosis, and eventual cirrhosis. The first line treatment currently used for PBC is ursodeoxycholic acid (UDCA) which works by enhancing choleresis. Most patients respond to first line treatment, however 30-40% will have an inadequate biochemical response to UDCA (so called UDCA non-responders), and worse long-term outcomes(1)(2). Patients with an adequate UDCA response are at lower risk of disease progression, whereas those with inadequate UDCA response are at higher risk. It is also widely supported that if the alkaline phosphatase (ALP) concentration is higher, the biliary injury caused by PBC is more severe. It is therefore shown that liver biochemistry during treatment with UDCA (the UDCA response) accurately predicts long-term outcome. Liver biochemistry, and in particular alkaline ALP is shown to be a surrogate marker of underlying disease activity.

Phenotypic stratification of PBC patients is vital to identify subgroups of patients with distinct disease trajectories. Identifying these subgroups will allow patients with the greatest need to be prioritised for newer treatment options, but also for these patients to be taken forward for molecular characterisation or clinical trials.

Latent class mixed modelling (LCMM)(3) is a type of latent class analysis used for phenotypic stratification. It is a semi-supervised statistical learning technique that uses longitudinal/repeated data as surrogate markers of disease activity, to identify distinct patient subgroups with specific disease trajectories.

We therefore applied LCMM on ALP data (as a surrogate marker of disease activity), prior to, and post treatment with UDCA, for phenotypic stratification in a PBC dataset with the aim to identify distinct patients subgroups with distinct disease trajectories.

## Materials & Methods

### Participants & Data capture

We considered two distinct datasets for our analysis. The discovery cohort consisted of 2,425 patients from 263 Health Trusts, collected by the UK-PBC Consortium. The validation cohort consisted of 3,765 patients from Health Trusts within Italy, as part of the Globe Consortium. In both datasets, participants were adults (≥18 years of age) with an established diagnosis of PBC, who were receiving first line treatment with UDCA. We considered retrospective longitudinal blood tests results and liver biochemistry tests from PBC patients before they began on UDCA and post treatment values. We selected patients with at least three serial blood results, but also with a minimum of two results post starting UDCA treatment. We included results from patients prior to their diagnosis of PBC, and initiation of therapy, up to the recorded end point or terminal end point as defined by the first recorded end-stage liver disease (ESLD) event. In the UK-PBC cohort ESLD was defined using four markers; variceal haemorrhage, ascites, liver transplantation and death related to liver disease. In the Globe cohort ESLD was defined using two markers: liver transplantation or death related to liver disease.

**Statistical analysis**

**Subject stratification from liver biochemistry trajectories**

We applied LCMM to longitudinal ALP measurements to identify subgroups of PBC patients with distinct disease trajectories (**see Supplementary materials**). We defined time in days, with time 0 as the starting point for UDCA treatment, and negative values representing time prior to UDCA treatment. We considered two models, either with a linear time trend (model 1) or a quadratic time trend (model 2). Each model was adjusted for an effect of treatment, and an interaction with time, including random effects of time. We used the Bayesian Information Criteria (BIC) (4), and the Akaike Information Criterion (AIC) (5) to optimise the number of latent classes (**see Supplementary methods**). Finally, we assessed the discrimination of the model to quantify how distinct the obtained latent classes were from each other.

**Clinical validation from liver biopsies**

We reviewed the liver biopsies from patients in the two extreme latent classes among the four identified by the latent class model, to look for underlying differences in disease activity. These biopsies were reviewed by a Consultant Histopathologist with expertise in liver disease, and independent of the data used in the latent class analysis.

We included biopsies taken at diagnosis, and during the patient disease course, but not biopsies taken at the time of transplantation. The Nakamuna scoring system was used to grade the biopsies, and to review any differences between the subgroups (6). This scoring system gives a score for; liver fibrosis (0-3), bile duct loss (0-3), and chronic cholestasis (0-3). These three scores are then added together to give an overall score (0-9) which is used to stage the PBC disease progression between stages 1-4. Stage 1 indicates no disease progression while Stage 4 indicates advanced progression. There is also a grading for the necroinflammatory activities of PBC based on cholangitis activity (CA) 0-3, and hepatitis activity (HA) 0-3. CA 0 indicative of no cholangitis activity, and CA 3 supportive of marked activity. HA 0 indicative of no interface hepatitis, and HA 3 supportive of marked interface hepatitis.

**Prediction of long-term ESLD from the ALP classification**

We used the patterns identified from the LCMM to review long-term patient outcomes in relation to ESLD, to assess the clinical meaning of the latent classes. We defined a composite event as the delay in days since the start of treatment to the first recorded marker of ESLD. Status 1 was recorded for patients with a marker of ESLD, and 0 for no marker. We defined a time-to-event, as the delay between time 0 and the first recorded event of ESLD. For subjects who did not experience ESLD, their time-to-event was censored at the end of the follow-up period. We used a Cox Proportional Hazards model to quantify the differences in the hazards rate of ESLD across the latent classes, and a Kaplan Meier curve to show survival probabilities for each of the classes over the course of the recorded data.

**Comparison with UK-PBC Risk Score in terms of predictive ability**

We compared how far the estimated latent classes would improve the predictions of ESLD when used in conjunction with the UK-PBC Risk Score(7) against the UK-PBC Risk Score alone. The UK-PBC Risk score was previously developed by our group to predict long-term outcome in PBC. It was derived from a Cox Proportional Hazards Model, and accurately predicts long-term outcomes in PBC at 5, 10, and 15 years respectively. The Risk Score is calculated from baseline blood results including albumin, and platelet count in conjunction with bilirubin, transaminases (ALT), and ALP after 12 months of UDCA treatment.

We reviewed individuals from the original model cohort with blood tests results available to complete the Risk Score. Using the model, each patient was assigned a latent class based on their ALP measurements collected up to 12 months after starting UDCA therapy. We first estimated a Cox model to describe the time to ESLD using the Risk Score alone, and a second model adjusted for the Risk Score in conjunction with the latent classes. The predictive ability of these two models was assessed using the area under the curve (AUC), and computed by internal 10-fold cross-validation.

## Results

### Participants

Our study analysed longitudinal ALP data from 1,601 participants (10.4% of patients were male, and 89.6% female) collected from the UK-PBC Consortium. The mean time interval was 1415 days after starting UDCA (**Table 1**), with 25,271 individual ALP results recorded. A total of 1,281 patients had an ALP result available prior to starting UDCA therapy. A plot of the ALP raw data is shown in **Figure 1**. Each trajectory on this plot indicates a patient's ALP results with time. Time 0 on the x axis indicates treatment start date with UDCA in days.

### Model selection

The linear and quadratic models were estimated with 1 to 5 latent classes. **Table 2** shows the log-likelihood, AIC, and BIC of the linear and quadratic models. The BIC and AIC for a quadratic trend was lower than for a linear model. This was irrespective of the number of latent classes within the model, indicating a better goodness of fit. Although the 5-class model had lower AIC, and BIC scores the numbers of patients in the 5th group was too low (18 subjects, 1.12% of the total cohort), and the fourth and fifth classes in this model predominantly consisted of the individuals in the fourth class in the 4-class model (**Table 3**). The 6-class model did not converge, likely due to the identification of a very small sub-group of participants. A quadratic 4- class model was therefore selected.

The selected model identified 4 classes of 576 (35.9%), 603 (37.7%), 145 (9.1%), and 277 (17.3%) subjects, respectively. **Figure 2** shows the estimated mean ALP trajectories of the four latent classes (thick lines), and the individual trajectories (thin lines), with the colour corresponding to the assigned latent class. Class 1 had a normal ALP before, and after UDCA treatment, Class 2 a rapid adequate ALP response to UDCA, Class 3 a rapid but persistently inadequate ALP response to UDCA, and Class 4 a markedly elevated ALP with no UDCA response.

Finally, the discrimination of the 4-class quadratic model was assessed in **Table 4**. The first row represents the mean posterior probability among subjects assigned to Class 1 to belong to Classes 1, 2, 3, or 4. We observed that the probabilities of misclassification were low, whilst the probability of good classification in Class 1 was 84.5%. **Table 4** also shows that the four classes obtained are adequately separated, with the diagonal elements higher than 73%. ALP is therefore shown (as a surrogate marker of disease activity and process) to be able to identify distinct subgroups in PBC.

**Confirmation of the classification based on ALP measurements on validation dataset**

After data cleaning, a total of 3,143 individuals from the Globe dataset were included in the validation cohort (**Table 5**). A total of 9.2% of patients were male, and 90.8% female. We applied the 4-class model to the Globe dataset to assign each subject to the class with the highest probability. The 4-class model showed good discrimination between the classes. This was evidenced by the diagonal elements in **Table 6** being higher than 68%.

We also used the model to predict the ALP results in each latent class, and compared the mean of ALP predictions to the mean ALP results weighted by the probability that participants belong to that class. **Figure 5** shows that these predictions were close to the weighted mean of the ALP observations, and lay within the $5^{th}$ and $95^{th}$ quantiles of the mean observations. This comparison between the predicted, and mean ALP results showed a good fit of the 4-class model to this external dataset.

**Clinical validation from liver biopsies**

We reviewed a subsection of patients from Classes 1 and 4 who had had liver biopsies completed during the course of their disease process. We analysed 11 liver biopsies in Class 1, and 7 in Class 4. Using the Nakanuma scoring system (6) the median stage score in Class 1 was 3, with a median HA of 1, and CA score of 1. In Class 4 the median stage score was 7, with a median HA of 1, and CA score of 3 (**Table 7**). Although the number of biopsies reviewed in each class was small there was a clear difference in the Stage score between the two groups (p = 0.014) and the CA (p = 0.035). The score in Class 4 was 7, supportive of Stage 4 disease, and advanced PBC progression. The score in Class 1 was 3, supportive of Stage 2 disease, and mild PBC progression. Class 4 also had marked cholangitis activity with a median CA of 3, indicative of chronic cholangitis in >2/3 of the portal tracts. This score compared to Class 1 where the median CA was 1, indicative of chronic cholangitis in < 1/3 of the portal tracts. The biopsies were therefore supportive of distinct phenotypic patient subgroups, with Class 4 having more advanced disease severity when compared to Class 1.

**Associations between latent classes and ESLD**

We assessed whether the latent classes had distinct outcomes in terms of ESLD. We reviewed the percentage of patients, in the UK-PBC cohort, with ESLD in each class. Classes 3 and 4 (20.7% and 31.1% respectively) had a higher percentage of patients with ESLD when compared to Classes 1 and 2 (5.2% and 16.3% respectively) (**Table 8**).

We also estimated the hazard ratio of time to ESLD for each of the classes compared to Class 1 (reference class). We used the Cox Proportional Hazards model adjusted for the latent class. Using this model the risk of ELSD for the patients in Class 4 was 7.63 times higher than for those in Class 1 (95% confidence interval (CI) (5.004-11.612)) (**Table 9**). We also used a Kaplan Meier curve to assess the probability to remain free of an ESLD recorded event against time in years for each of the classes. Class 4 not only had a more rapid decline in survival probability over time, but also a lower survival probability at each time point over the course of the data recorded (when compared to all other classes, and in particular Class 1) (**Figure 5**). For example, patients in Class 4 had a 50% survival probability at 15 years post UDCA treatment, as compared to Class 1 who had a 87.5% survival probability.

We also reviewed the number of patients with ESLD in the validation cohort. In this cohort two markers were recorded, which were liver transplantation, or death related to ESLD. We showed that Class 4 had a higher percentage of patients with ESLD when compared to the other classes, and in particular Class 1 (14.3% compared to 1.1%) (**Table 10**).

Therefore, the phenotypic classification achieved by the model using ALP measurements was shown to be meaningful in terms of disease severity. The latent classes showed different risks of developing ESLD, and differing probabilities to be free of ESLD.

**Comparison with UK-PBC Risk Score**

We used a total of 1,118 individuals from the discovery cohort to compute the UK-PBC risk score. These patients had baseline tests of; albumin, platelet count, and bilirubin, recorded together with ALT, and ALP at 12 months post UDCA commencement. We used a Cox model to predict ESLD from, firstly the UK-PBC Risk Score alone or, secondly the UK-PBC Risk Score combined with individual estimated latent classes. Using these two models the probability of ESLD within a 5-year period, and the AUC was computed. The AUC score for the UK-PBC risk score was 0.82 (95% CI 0.76-0.87), and the AUC with the addition of the classes was 0.81 (95% CI 0.76-0.87). The AUC score showed that the latent class model did not improve the UK-PBC Risk Score.

**Discussion**

The best model for ALP measurements identified 4 well discriminated groups, each with specific disease trajectories. Class 1 had a normal ALP before, and after treatment with UDCA. Class 2 had a rapid, adequate response to treatment. Class 3 had a rapid but persistently inadequate UDCA response, and Class 4 a markedly elevated ALP with no UDCA response. Class 4 was also shown (using a Cox Proportional Hazards model for the time to ESLD) to have a higher risk of ESLD in comparison to Class 1.

We validated the model using an external patient cohort from the Globe Consortium. This cohort did have a lower number of patients with ESLD events recorded compared to the UK-PBC dataset. The Globe cohort had 6.2% of patients with an ESLD event recorded, whilst the UK-PBC dataset had 15.2%. This may have been due to a variety of reasons including; only two markers used to define ESLD in the Globe cohort compared to the UK-PBC cohort's four markers, a reduced period of follow-up in the Globe cohort with fewer ESLD events recorded, and a higher proportion of patients in the UK-PBC cohort being recruited from tertiary liver transplant centres, thereby with higher rates of ESLD events recorded. Allowing for the differences in ESLD recording between the two cohorts, the goodness-of-fit of the ALP predictions of the 4-latent class model on this cohort confirmed its portability to external ALP data. The model was also able to categorise patients into 4 classes with differing risks to develop ESLD in the Globe dataset.

We were limited by the number of liver biopsies within the study, as the majority of patients are diagnosed with PBC without the need for this invasive procedure, and it is not required in the PBC diagnostic classification system(8). Even with limited liver biopsies we found that Class 4 had histological signs of more advanced disease progression and cholangitis activity when compared to patients in Class 1. This was supportive of the higher risk of disease progression and severity leading to ESLD as shown in Class 4 compared to Class 1 (**Table 8**).

ALP, the surrogate marker of disease activity which has been traditionally used to monitor patients, is shown to be a reliable biochemical result to monitor disease activity over time and to discriminate between patient groups with increased probability of ESLD. This modelling approach suggests that there are discrete patterns of UDCA response, likely reflecting distinct mechanisms of disease, and that patients with persistent disease activity have a worse outcome. It also shows that the current step-up approach to treatment of PBC disadvantages those in the greatest need, particularly patients in Class 4, who are most likely to have an inadequate UDCA response and wait the longest to achieve effective treatment.

Although the latent class model did not improve the predictive ability for ESLD predictions compared to the UK-PBC Risk score, this model can be used to inform future work. This future work could focus on exploring the mechanistic differences between subgroups with the aim to identifying the processes behind them, such as Class 1 (UDCA responders) compared to Class 4 (UDCA non-responders).

This study has shown that computational models can be used for phenotypic stratification of patients, identifying groups with different PBC disease trajectories and distinct long-term outcomes. It also shows that a conventional surrogate marker of disease activity, ALP, can identify these distinct phenotypic subgroups. Identifying these disease patterns can inform future studies aimed at understanding the mechanisms behind these subgroups, which in turn will lead to newer treatment options, and improved outcomes, in particular for those patients with the highest risk of ESLD.

al. The British Society of Gastroenterology/UK-PBC primary biliary cholangitis treatment and management guidelines. Gut. 2018;67(9):1568–94.

**Figure Legends**

**Figure 1.** Individual ALP longitudinal results over time in days. Each trajectory shows one patient's ALP results with time. Time <0 indicates UDCA starting. Time >0 indicates ALP results after UDCA starting.

**Figure 2.** Individual ALP trajectories (thin lines) and estimated class- specific ALP trajectories (thick lines), as functions of time in days. Individual patient trajectories are depicted using the same colour as their assigned classes.

**Figure 3.** Class- specific trajectories obtained by latent class mixed model for ALP over time.

**Figure 4.** Class-specific Kaplan Meier curves showing survival probabilities against time in years for each of the latent classes. The dashed lines represent the 95% confidence intervals of the survival curves.

**Figure 5.** Goodness-of-fit visualisation of the 4-class model on the Globe dataset. Solid lines represent the weighted mean of the ALP observations at each 6 month time interval. Crosses represent the weighted mean of the ALP predictions at each 6 month time interval. Dashed lines represent the 5$^{th}$ and 95$^{th}$ quantiles of the ALP observations.

# Main Figures

Individual patient ALP trajectories over time

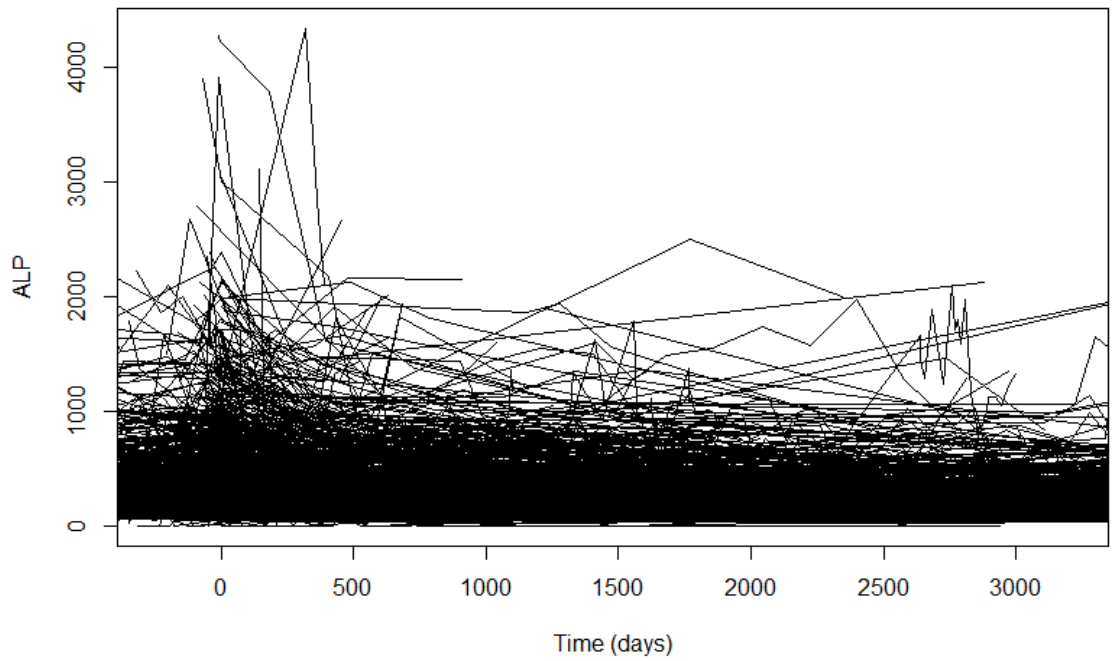

**Figure 1.** Individual ALP longitudinal results over time in days. Each trajectory shows one patient's ALP results with time. Time <0 indicates UDCA starting. Time >0 indicates ALP results after UDCA starting.

Individual patient ALP trajectories and estimated class specific ALP trajectories with time

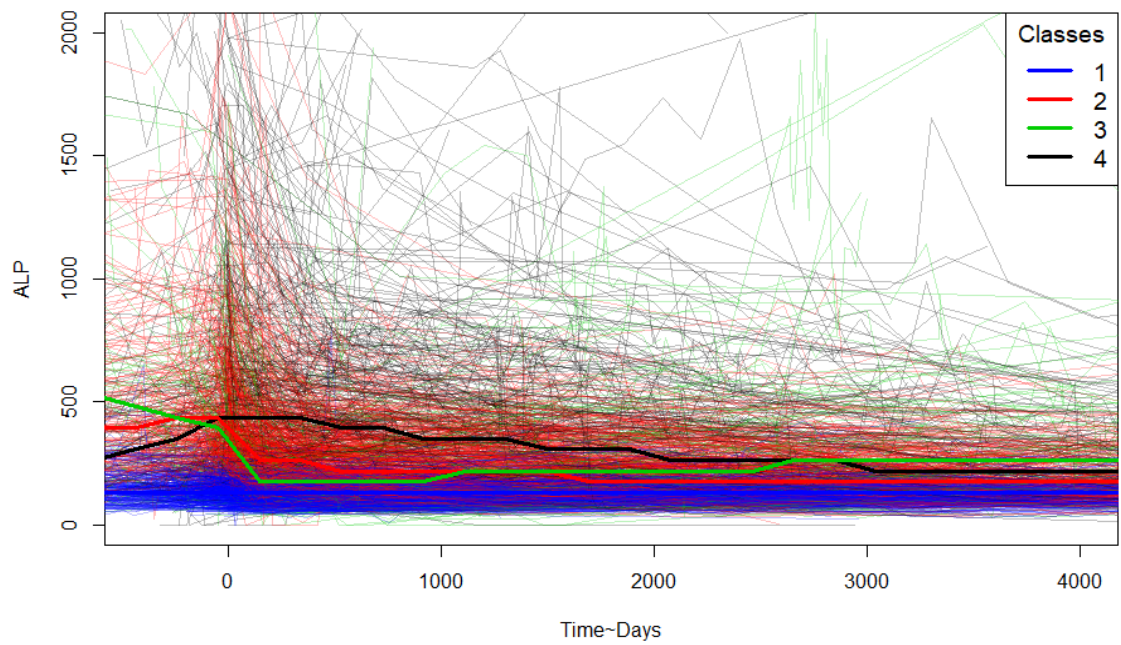

**Figure 2.** Individual ALP trajectories (thin lines) and estimated class- specific ALP trajectories (thick lines), as functions of time in days. Individual patient trajectories are depicted using the same colour as their assigned classes.

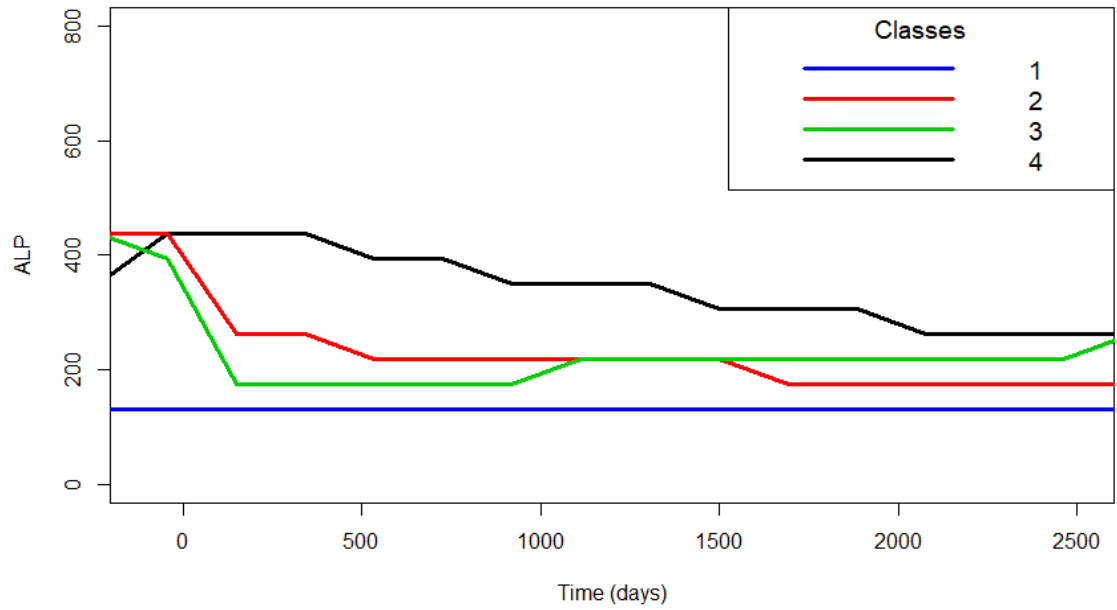

**Figure 3.** Class- specific trajectories obtained by latent class mixed model for ALP over time.

Kaplan Meir curve showing survival probabilities for each of the latent classes

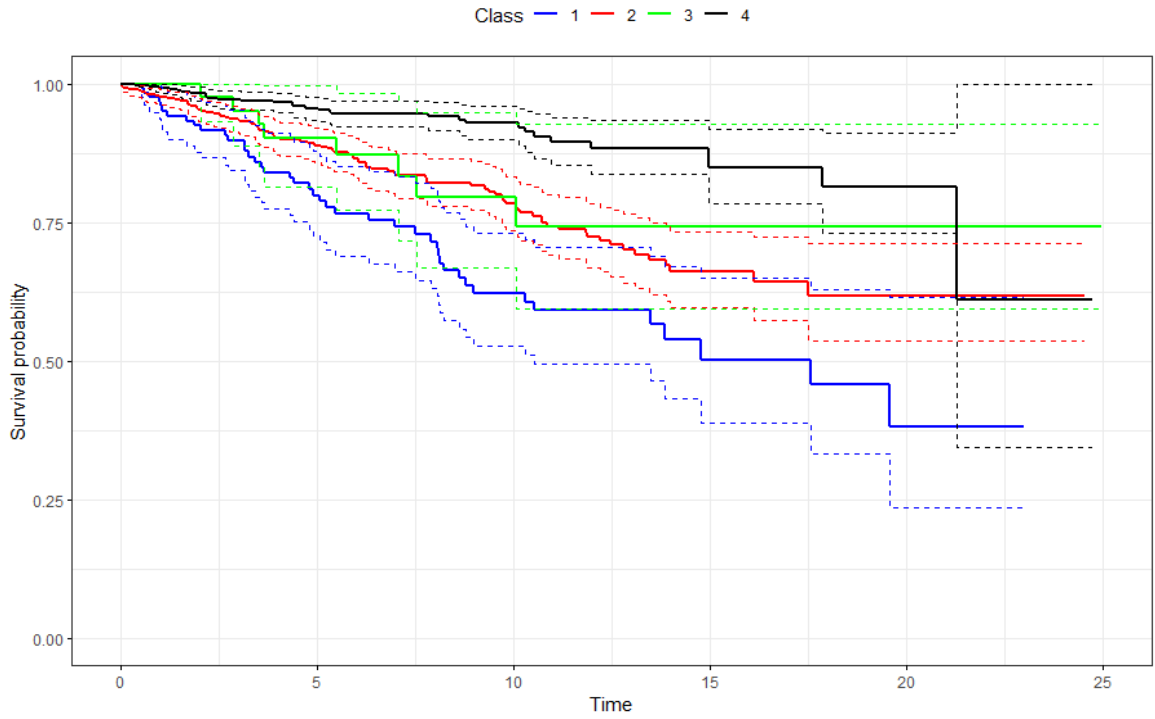

**Figure 4.** Class-specific Kaplan Meier curves showing survival probabilities against time in years for each of the latent classes. The dashed lines represent the 95% confidence intervals of the survival curves.

Goodness of fit visualisation for the 4 class model using the Globe dataset

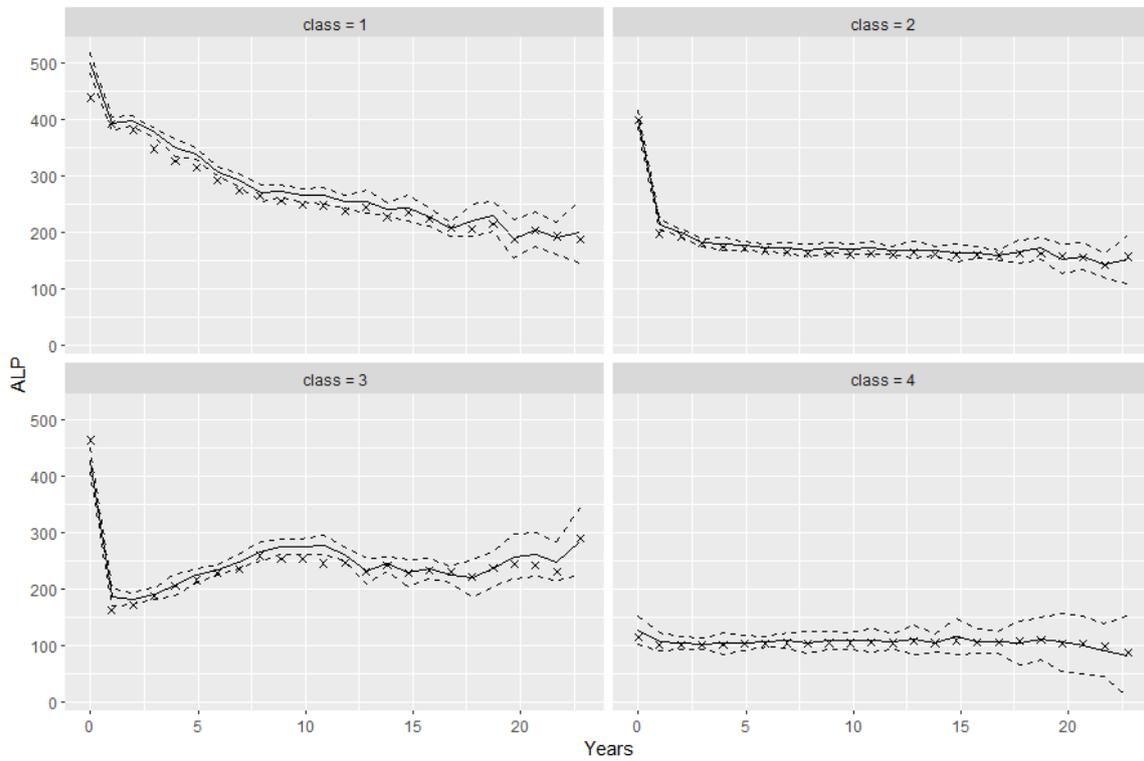

**Figure 5.** Goodness-of-fit visualisation of the 4-class model on the Globe dataset. Solid lines represent the weighted mean of the ALP observations at each 6 month time interval. Crosses represent the weighted mean of the ALP predictions at each 6 month time interval. Dashed lines represent the 5$^{th}$ and 95$^{th}$ quantiles of the ALP observations.

**Table 1. Summary of ALP results for patients used for classification**

| No. of patients | Blood test | Max no. results per patient | Min no. results per patient | Mean no. results per patient | Max ALP result recorded | Min ALP result recorded | Mean ALP result recorded | Median result recorded | Min time interval (days) | Max time interval (days) | Mean age |
|---|---|---|---|---|---|---|---|---|---|---|---|
| 1601 | ALP | 99 | 3 | 15.78 | 4335 | 0 | 239 | 161 | -8536 | 10567 | 55.74 |

Summary of patient cohort longitudinal ALP blood tests results used in latent class mixed model. ALP: alkaline phosphatase

**Table 2. Summary of goodness of fit results models with 1 to 5 classes**

| Fit Statistics | 1-class model | 2-class model | 3-class model | 4-class model | 5-class model |
|---|---|---|---|---|---|
| Log-likelihood | -141947.92 | -141436.5 | -141047.22 | -140857.44 | -140583.25 |
| AIC | 283925.84 | 282918.99 | 282156.43 | 281792.88 | 281260.5 |
| BIC | 284006.51 | 283042.69 | 282323.16 | 282002.64 | 281513.28 |

Results of latent class analysis using a quadratic time trend model. AIC: Akaike Information Criterion, BIC: Bayesian Information Criterion

**Table 3. Summary of number of patients within each group using 1 to 5 class models.**

| Latent class model | Number of total patients in each class ||||| 
|---|---|---|---|---|---|
| | 1 | 2 | 3 | 4 | 5 |
| 2 | 260 | 1341 | n/a | n/a | n/a |
| 3 | 261 | 1313 | 27 | n/a | n/a |
| 4 | 576 | 603 | 145 | 277 | n/a |
| 5 | 167 | 541 | 608 | 267 | 18 |

Number of patients within each latent class identified using 1 to 5 latent class mixed model using the quadratic time trend model.

**Table 4. Discrimination table**

| Assigned Class | Probability to belong to class 1 | Probability to belong to class 2 | Probability to belong to class 3 | Probability to belong to class 4 |
|---|---|---|---|---|
| 1 | 0.8453 | 0.0950 | 0.0208 | 0.0389 |
| 2 | 0.1064 | 0.7352 | 0.0954 | 0.0630 |
| 3 | 0.0328 | 0.1307 | 0.8260 | 0.0106 |
| 4 | 0.0290 | 0.1606 | 0.0364 | 0.7741 |

Mean posterior probabilities for patients to belong to each class within the four latent class model.

**Table 5. Summary of ALP results for Globe patients used in the validation cohort.**

| No. of patients | Blood test | Max no. results per patient | Min no. results per patient | Mean no. results per patient | Max ALP result recorded | Min ALP result recorded | Mean ALP result recorded | Median ALP result recorded | Min time interval (days) | Max time interval (days) | Mean age |
|---|---|---|---|---|---|---|---|---|---|---|---|
| 3143 | ALP | 42 | 3 | 10.4 | 2508 | 21 | 356.8 | 262.0 | 0 | 6300 | 52.3 |

Summary of Globe Patient Cohort result statistics for longitudinal ALP blood results.

**Table 6. Summary of posterior probabilities for patients to belong to each class within the model using the validation cohort.**

| Assigned Class | Probability to belong to class 1 | Probability to belong to class 2 | Probability to belong to class 3 | Probability to belong to class 4 |
|---|---|---|---|---|
| 1 | 0.7480 | 0.1803 | 0.0557 | 0.0160 |
| 2 | 0.1293 | 0.6855 | 0.1198 | 0.0654 |
| 3 | 0.0764 | 0.2038 | 0.7005 | 0.0193 |
| 4 | 0.0715 | 0.1643 | 0.0373 | 0.7269 |

Mean posterior probabilities to belong to each class, given the assigned class, within the four latent class model using the Globe Patient Cohort.

**Table 7. Summary of histological scoring for liver biopsies between class 1 and 4**

| Nakanuma scoring criteria | Class 1 (n= 11) | Class 4 (n = 7) | p |
|---|---|---|---|
| Score | 7 (0-9) | 3 (5-9) | 0.0135 |
| Cholangitis activity | 1 (0-3) | 3 (1-3) | 0.035 |
| Hepatitis activity | 0 (0-3) | 1 (0-2) | 0.103 |

Nakanuma histological staging and grading system for PBC in a latent class 1 and 4. Data shown in the parentheses indicate the median result. Statistical significance was calculated using the Mann- Whitney test (2 tailed) for continuous variables. p: value comparing class 1 and class 4 results.

**Table 8. Summary of ESLD results for each patient within each class in discovery cohort.**

| Latent Class | Number of ESLD=1 | Percentage of ESLD within class |
|---|---|---|
| 1 | 30 | 5.21 |
| 2 | 98 | 16.25 |
| 3 | 30 | 20.69 |
| 4 | 86 | 31.05 |

Number of patients within each class who developed an end stage liver disease event during the period of data collection. ESLD: end stage liver disease

**Table 9. Summary of hazard ratio analysis for the four latent classes.**

| Latent Class | Regression coefficient | Hazard ratio | Standard error of coefficient | Wald statistic | p value |
|---|---|---|---|---|---|
| 1 | n/a | n/a | n/a | n/a | n/a |
| 2 | 1.2460 | 3.4764 | 0.2114 | 5.893 | $3.79e^{-9}$ |
| 3 | 1.5095 | 4.5244 | 0.2606 | 5.792 | $6.95e^{-9}$ |
| 4 | 2.0315 | 7.6255 | 0.2150 | 9.450 | $<2e^{-16}$ |

Estimates of the Cox Proportional Hazards model adjusted for the four classes identified by the quadratic latent class model.

**Table 10. Summary of ESLD results for each patient within each class within the validation cohort.**

| Latent Class | Total number of individuals | Percentage of total cohort | Number of ESLD=1 | Percentage of ESLD within class |
|---|---|---|---|---|
| 1 | 539 | 17.1 | 6 | 1.1 |
| 2 | 1602 | 51.0 | 66 | 4.1 |
| 3 | 241 | 7.7 | 14 | 5.8 |
| 4 | 761 | 24.2 | 109 | 14.3 |

Number of patients within each class within the Globe Patient Cohort who developed an end stage liver disease event during the period of data collection. ESLD: end stage liver disease.